\documentclass[sigconf,nonacm]{acmart}
\renewcommand\footnotetextcopyrightpermission[1]{}
\usepackage{amsmath}
\usepackage{amsfonts}
\usepackage{booktabs}
\usepackage{tabularx}
\usepackage{array}
\newcolumntype{L}[1]{>{\raggedright\arraybackslash}p{#1}}
\newcolumntype{Y}{>{\raggedright\arraybackslash}X}
\usepackage{microtype}
\makeatletter
\def\@secfont{\bfseries\Large\section@raggedright}
\makeatother
\title{Evidence Coverage for Intent-Bound Execution:\\
Scope, Obligations, and Cutoff Reasoning}

\author{Mengting Wu}
\authornote{Corresponding author.}
\email{chloe@havenlon.com}
\affiliation{%
  \institution{Chengdu Havenlon Security Technology Co., Ltd.}
  \city{Chengdu}
  \country{China}
}

\author{Lin Wang}
\affiliation{%
  \institution{Chengdu Havenlon Security Technology Co., Ltd.}
  \city{Chengdu}
  \country{China}
}

\author{Yong Zhang}
\affiliation{%
  \institution{Chengdu Havenlon Security Technology Co., Ltd.}
  \city{Chengdu}
  \country{China}
}

\author{Jiang Deng}
\affiliation{%
  \institution{Chengdu Havenlon Security Technology Co., Ltd.}
  \city{Chengdu}
  \country{China}
}

\begin{document}
\begin{abstract}
A verifier may authenticate every available record and still lack grounds to call an execution account complete. Such a claim requires a justified account of which records were due for the execution being assessed. We present an analytical model for retrospective coverage of declared execution-evidence obligations. Its scope binds a structured Intent, an exact Candidate, a selected analytical attempt, an execution and evidence boundary, a stage horizon, a fixed record-obligation profile, a named verifier, and an assessment cutoff. Branch and trigger premises determine obligation instances; source competence, content, integrity, and object and stage bindings determine admissibility. We distinguish closure of the obligation inventory from closure of the relevant verifier view, and define three reporting results: \mbox{\texttt{COMPLETE\_WITHIN\_SCOPE}}, \texttt{INCOMPLETE}, and \texttt{UNKNOWN}. These results concern current coverage of obligations due at the cutoff. Execution progress, external outcome knowledge, and historical delivery timeliness are reported separately. Constructed service-principal-disablement cases demonstrate complete dispatch and refusal branches, a due but missing final-result record, subsequent coverage after late delivery, and the limits of extending one selected attempt's coverage to all attempts. The contribution is an execution-specific composition of scope, branch, horizon, obligations, admissibility, view, and cutoff. Completeness remains conditional on the declared profile and assessment premises.

\end{abstract}
\maketitle
\raggedbottom
\hypersetup{
  pdftitle={Evidence Coverage for Intent-Bound Execution: Scope, Obligations, and Cutoff Reasoning},
  pdfauthor={Mengting Wu; Lin Wang; Yong Zhang; Jiang Deng},
  pdfsubject={Evidence coverage for intent-bound execution}
}
\ifXeTeX
  
\fi
\section{Introduction}
\label{sec:1}

An execution account can contain individually authentic records while omitting a record that its reader should have received. Conversely, the absence of a later-stage record need not indicate an omission: an attempt that ended in a justified terminal refusal may have no dispatch, invocation, or final-result record to supply. Both observations concern the set of records required for the execution being assessed. Verifying the records already retrieved does not by itself identify that set.

Secure logging, signed message streams, transparency logs, and provenance provide foundations for this analysis. They protect generated records, expose some missing stream positions, establish inclusion and consistency for submitted objects, or represent and validate recorded relationships \cite{schneierKelsey1999,rfc5848,rfc9162,provDM,provConstraints,souilah2009,chan2020}. Provenance research also explicitly examines completeness and missing records \cite{souilah2009,chan2020}. Those properties can contribute to an execution account. An additional application-specific specification is needed to identify the branch-specific obligations that should be visible to a particular verifier at a particular cutoff. Likewise, a pre-execution profile can determine which evidence a Candidate must present to obtain execution authority \cite{p2}, while a retrospective assessment still needs to specify records of the stages actually triggered.

Our research question is:

\noindent
\textbf{What does a verifier's evidence view justify about coverage of declared execution-evidence obligations for a stated execution scope, branch, and assessment cutoff?}

We answer this question with a compact analytical model and constructed cases. A record-obligation profile is declared before assessment. It generates instances from identified triggers within a stated stage horizon. A positive conclusion requires grounds for the branch and for the completeness of this inventory, as well as admissible evidence for every applicable obligation due at the cutoff. A supported specific gap requires grounds that the relevant verifier view contains no admissible record satisfying that obligation. Otherwise, coverage remains unresolved.

Branch sensitivity is central to the execution-specific account. Under the constructed profile, a terminal refusal requires Intent, Candidate, adjudication, and refusal records. Later-stage obligations are inapplicable because their triggers did not occur under that justified branch. A dispatch branch with a known invocation instead activates status and final-result reporting obligations. The assessment therefore operates over generated obligation instances, with a fixed profile defining both branches.

The paper makes three contributions:

\begin{enumerate}
\item \textbf{An execution-scoped evidence-coverage model.} We define the object of assessment by binding Intent, exact Candidate, selected analytical attempt, boundary, stage horizon, fixed profile, verifier, and cutoff. We distinguish the profile from the inventory it generates for that scope.
\item \textbf{A branch- and cutoff-sensitive assessment method.} We specify applicability, due times, admissibility, obligation-inventory closure, and verifier-view closure, with explicit precedence for \texttt{INCOMPLETE}, \mbox{\texttt{COMPLETE\_WITHIN\_SCOPE}}, and \texttt{UNKNOWN}. The report separates coverage from progress, outcome knowledge, and historical delivery timeliness.
\item \textbf{Analytical demonstrations.} Constructed cases show complete dispatch and refusal branches, a due but missing final-result record, later delivery that closes current coverage while preserving an earlier deadline miss, and complete selected-attempt coverage with unresolved all-attempt coverage.
\end{enumerate}

The model evaluates retrospective record coverage. It uses action-relative scope ideas from the authority analysis in \cite{p1} and structured Intent and Candidate ideas from \cite{p2}. Its reporting results carry no execution permission. We make no claim to a new cryptographic mechanism, a general missing-record detector, or universal discovery of execution obligations.

\section{Related Work and Positioning}
\label{sec:2}

\subsection{Logging and transparency}
\label{sec:2.1}

Schneier and Kelsey protect audit entries generated before compromise of a logging machine, limiting disclosure and making specified modifications or destruction detectable under their protocol assumptions \cite{schneierKelsey1999}. This is a protection property for generated history. For an execution assessment, the required event-to-record mapping and its branch-specific inventory must additionally be supplied.

RFC 5848 adds authentication, integrity, replay resistance, sequencing, and detection of missing transmitted syslog messages. Signature Block information supports delivery and ordering checks under the stream's assumptions (§7; §§8.5--8.7) \cite{rfc5848}. It can help establish a record range or a transport gap; it does not, by itself, specify which execution stages must generate messages.

RFC 9162 defines Merkle inclusion and append-only consistency proofs (§§2.1.3--2.1.4). Its Signed Certificate Timestamp commits the log to inclusion within a Maximum Merge Delay (§4; §§4.1, 4.8) \cite{rfc9162}. Inclusion deadlines and delayed inclusion are therefore existing concepts. Our question additionally identifies which execution records ought to be available in a named verifier view. A transparency log can carry such records when an application supplies submission and obligation rules.

\subsection{Provenance and completeness}
\label{sec:2.2}

PROV-DM supplies entities, activities, agents, and relationships such as generation, usage, and derivation, with extension points for application-specific information \cite{provDM}. PROV-Constraints supplies uniqueness, ordering, impossibility, and type constraints for valid provenance instances \cite{provConstraints}. Representation and instance validity can support the bindings assessed here; they do not alone enumerate the execution-record obligations of an independently specified profile.

Souilah, Francalanza, and Sassone model distributed provenance using an annotated calculus. Their correctness and completeness definitions use a monitored global log as a proof device (§§3.3--3.5). They show that their strong completeness condition is not preserved and argue for provenance adequate to its intended application (§5) \cite{souilah2009}. Our declared scope and profile similarly make adequacy application-relative, but the assessed object is a due execution-record obligation in a verifier view.

Chan et al. explicitly investigate missing and irregular provenance records. They compare generated graphs with a control graph assumed to describe known activities correctly and completely, and demonstrate the approach through manual experiments (§§3--4) \cite{chan2020}. Our method also requires a justified basis for what should be present. It expresses that basis as branch- and trigger-generated obligations, with source and object bindings and separate current-coverage and delivery-time findings.

\subsection{Prior execution analyses}
\label{sec:2.3}

The authority-decomposition framework in \cite{p1} asks which trust-domain coalitions can realize a designated protected transition and distinguishes causal control from control over its authoritative evidentiary account. We use its action-relative boundary perspective. We do not derive realization witnesses, coalitions, bypass, veto coverage, or independence from a record inventory.

The execution-boundary conformance profile in \cite{p2} specifies Candidate-bound adjudication and release through an Execution Release Contract, a separate Execution Grant, and Redemption-time conditions. Its Evidence Obligations concern evidence required for an execution-authority decision. The obligations defined here concern records required for retrospective reporting of triggered stages. Neither inventory nor result label replaces the policies, ERC, Grant lifecycle, or successful Redemption semantics of \cite{p2}.

\subsection{Comparison and residual contribution}
\label{sec:2.4}

\begin{table*}[t]
\caption{Related abstractions and the additional execution-coverage specification. Guarantees remain subject to each cited work's assumptions.}
\label{tab:related}
\centering
\fontsize{8}{9.6}\selectfont
\setlength{\tabcolsep}{2pt}
\renewcommand{\arraystretch}{1.12}
\begin{tabularx}{\textwidth}{@{}L{0.20\textwidth} L{0.20\textwidth} Y L{0.23\textwidth}@{}}
\toprule
\textbf{Research family} & \textbf{Primary object} & \textbf{Typical guarantee or analytical property} & \textbf{What remains for this paper} \\
\midrule
Secure audit logging \cite{schneierKelsey1999} & Generated audit entries & Protection of pre-compromise history against specified disclosure and undetected manipulation & Declare required execution records, their triggers, and their scope \\
\addlinespace[3pt]
Signed and sequenced logging \cite{rfc5848} & Transmitted messages and Signature Blocks & Authentication, sequencing, and detection of missing messages within the supported stream & Map stream objects to branch-specific execution obligations \\
\addlinespace[3pt]
Transparency logging \cite{rfc9162} & Submitted objects and log commitments & Inclusion, consistency, and deadline-bound inclusion promises & Specify execution obligations and availability in the named verifier view \\
\addlinespace[3pt]
Provenance models and validation \cite{provDM,provConstraints} & Entities, activities, agents, and relations & Representation and consistency checks over provenance instances & Supply the execution-specific inventory and assessment cutoff \\
\addlinespace[3pt]
Provenance completeness and quality checking \cite{souilah2009,chan2020} & Annotated histories or generated graphs relative to a reference basis & Formal completeness analysis and identification of missing or irregular provenance & Compose scope, branch, horizon, source/\allowbreak{}binding predicates, view closure, and due times \\
\addlinespace[3pt]
Authority decomposition \cite{p1} & Action-relative causal realization structures & Sufficient coalitions and evidentiary-control distinctions & Assess retrospective records without deriving causal authority \\
\addlinespace[3pt]
Execution-boundary conformance \cite{p2} & Candidate, adjudication, ERC, Grant, and Redemption & Conditions for receiving and redeeming execution authority & Assess triggered-stage records without issuing execution permission \\
\bottomrule
\end{tabularx}
\end{table*}

The residual contribution is the composition of \textbf{scope + branch + stage horizon + obligation inventory + admissible source/\allowbreak{}binding + verifier view + cutoff} into a bounded coverage judgment. Integrity versus completeness, missing-record detection, and timeliness are foundations of this composition. Our method can be instantiated using the cited logging or provenance mechanisms; it does not claim that they cannot represent the required records.

\section{Execution-Evidence Coverage Model}
\label{sec:3}

\subsection{Declared scope}
\label{sec:3.1}

We define the manuscript-local assessment scope

\[
S=\langle I,C,A,B,H,P,W,t_c\rangle.
\]

\begin{center}
\fontsize{8}{9.6}\selectfont
\setlength{\tabcolsep}{2pt}
\renewcommand{\arraystretch}{1.10}
\begin{tabularx}{\columnwidth}{@{}L{0.19\columnwidth} Y@{}}
\toprule
\textbf{Component} & \textbf{Meaning in this paper} \\
\midrule
\(I\) & A fixed structured Intent identifying the requested action and relevant objects \\
\addlinespace[3pt]
\(C\) & One exact materialized Candidate associated with \(I\), including the parameters needed to distinguish the action under assessment \\
\addlinespace[3pt]
\(A\) & A selected analytical handle for the attempt being examined, with an explicit basis connecting it to \(I,C\) \\
\addlinespace[3pt]
\(B\) & The declared execution and evidence boundary: included components, interfaces, stage observations, and collection limits \\
\addlinespace[3pt]
\(H\) & The declared stage horizon, including any branch-conditioned terminal stages; it is not shortened merely because later records are missing \\
\addlinespace[3pt]
\(P\) & A fixed record-obligation profile defining branch guards, triggers, instance generation, bindings, admissible sources, deadlines, and record predicates \\
\addlinespace[3pt]
\(W\) & The named verifier whose available evidence is assessed \\
\addlinespace[3pt]
\(t_c\) & The cutoff for this assessment \\
\bottomrule
\end{tabularx}
\end{center}

The handle \(A\) indexes an analytical selection. It is not a new execution-protocol identifier, a claim that one Intent has only one attempt, or a redefinition of the boundary-relative attempt in \cite{p1}. Its identity and selection basis are premises of the assessment.

Associated assessment premises document how the branch, triggers, source qualifications, collection boundaries, and time comparisons are justified. The tuple identifies the claim; these premises support it. The tuple and the definitions below are proposed here and are not assumed to be an existing project-wide schema.

Let \(V_W(t)\) denote the records available for inspection in \(W\)'s relevant view at time \(t\), with their associated receipt information. A retrieval may expose only a subset of this view. Membership at an earlier cutoff must be supported for that cutoff; a later retrieval cannot simply be treated as the earlier view.

\subsection{Obligation instances and admissibility}
\label{sec:3.2}

We define a record-obligation instance

\[
\begin{aligned}
q=\langle &\mathrm{trigger},\mathrm{branch},\mathrm{bindings},\\
&\mathrm{sources},d_q,\mathrm{predicate}\rangle.
\end{aligned}
\]

The trigger identifies the event or condition that activates this instance. The branch is its applicability guard. The bindings specify the relevant Intent, Candidate, selected attempt, request, and predecessor relationships. The sources specify the competent attesters and their qualification conditions. The deadline \(d_q\) specifies availability at \(W\). The predicate specifies required content and profile-specific integrity, stage, and time checks. The stage to which an instance belongs is fixed by \(P\), so it can be tested against \(H\).

An instance exists independently of whether its requested record was produced. For example, an invocation may activate a final-result reporting duty; the duty is not postponed until a final result or its record exists. Distinct triggers can generate distinct instances of the same type. Seeing one invocation record does not cover an additional invocation obligation.

We write \(\mathrm{Applicable}(q,S)\) when the trigger is established for the selected scope, the justified branch admits the instance, and its stage is within \(H\). We write

\[
\mathrm{Due}(q,t_c)\iff d_q\leq t_c.
\]

The deadline is inclusive: delivery at \(d_q\) counts as delivery by the deadline. If a trigger, branch, or deadline basis is unresolved, the corresponding assessment remains unresolved rather than being treated as false.

\(\mathrm{Admissible}(r,q)\) means that \(r\) meets the declared source competence, content, integrity, identity-binding, causal/stage, and any independently specified time predicates for \(q\). Binding can be direct or through verified correlations, such as an external request identifier linked to the exact Candidate in an invocation record. A source need not attest stages outside its designated role.

Admissibility is a rule for accepting a record for this duty. Authentication alone does not establish source competence, and admissibility does not establish the external truth of the assertion. A delivery deadline is assessed separately; late arrival alone does not make a record inadmissible for current coverage. Any additional freshness or event-time requirement must be independently declared in \(P\).

The underlying coverage relation is

\[
\begin{aligned}
&\mathrm{Covered}(q,V_W(t_c))\\
&\quad\iff\exists r\in V_W(t_c):\mathrm{Admissible}(r,q).
\end{aligned}
\]

An assessor may know this existential condition to hold, establish that no satisfying record exists in the relevant view, or leave it unresolved. An unresolved candidate check is not a negative finding. One definitely admissible record can establish coverage even if another candidate remains unresolved.

\subsection{Time and the supporting record properties}
\label{sec:3.3}

We keep four time concepts separate:

\begin{itemize}
\item \textbf{Event time:} when the source asserts or observes the described event.
\item \textbf{Record creation time:} when the record is produced, where relevant to its predicates.
\item \textbf{First verifier receipt time:} when \(W\) first obtains that record, under the stated receipt-history basis.
\item \textbf{Assessment cutoff:} the time defining the view and due obligations being assessed.
\end{itemize}

Deadlines concern availability at \(W\). Source event and creation timestamps cannot substitute for \(W\)'s receipt history. Comparing these times requires a declared clock or conversion basis; the constructed cases stipulate one common observer clock.

Record integrity, observed continuity, and justified record-range completeness can support this assessment. Integrity concerns the authenticity and protection of records; continuity concerns declared predecessor or sequence relations in an observed segment; range completeness additionally needs a justified stream identity and endpoint. Execution-obligation coverage concerns the applicable due instances generated by \(P\). These supporting properties are useful checks, not a separate novelty claim.

\section{Coverage Assessment Semantics}
\label{sec:4}

\subsection{Obligation-inventory closure}
\label{sec:4.1}

For a branch \(b\) declared by \(P\), we write \(\mathrm{BranchJustified}(b,S)\) when admissible branch-decision records, appropriately scoped terminal disposition evidence, or an explicit analytical premise in a constructed example support selecting \(b\) for \(S\). The basis must identify the relevant scope and any terminal premise used to exclude later triggers. Missing downstream records do not supply that basis: absence of an invocation record alone does not justify terminal refusal. Unresolved or conflicting branch evidence remains unresolved.

A record may both support the factual basis for a branch and satisfy a reporting obligation about it under the declared predicates. This does not permit retrospective branch selection from whichever received record set yields the smallest inventory.

Let \(Q_S\) be the inventory of instances generated by \(P\) from triggers within \(B,H\) for the selected attempt, under the justified branch, up to \(t_c\). It includes triggered instances whose deadlines have not yet arrived. Future, untriggered stage schemas remain in \(P\); they are not silently counted as completed.

We write \(\mathrm{ClosedObligations}(S)\) when there is sufficient basis to justify that \(Q_S\) is the complete inventory for this declared profile, scope, branch, horizon, and cutoff. This requires:

\begin{enumerate}
\item a fixed \(P,B,H\) and selection basis for \(I,C,A\);
\item \(\mathrm{BranchJustified}(b,S)\), including any terminal-branch premise used to exclude later triggers;
\item adequate trigger and instance-enumeration information to rule out an omitted applicable instance within this scope.
\end{enumerate}

This closure is relative to the profile and stated premises. It does not assert that \(P\) specifies every record that any conceivable system should require. A ledger created from received records is not, by itself, a closed obligation inventory. Nor may a missing later record be used to infer that its trigger did not occur.

If branch or trigger enumeration is unresolved, the assessor retains a provisional inventory and its unresolved parts. Common obligations may still be known to apply across possible branches. Thus an unresolved overall inventory can coexist with a supported specific gap; it prevents a positive completeness result but need not prevent \texttt{INCOMPLETE}.

\subsection{Verifier-view closure and absence}
\label{sec:4.2}

We write \(\mathrm{ClosedView}(q,W,t_c)\) when the assessment has sufficient basis that inspection includes every record in \(W\)'s relevant view that could satisfy \(q\). The basis must identify the relevant scope, collection interfaces, retention interval, and indexing or retrieval coverage, including alternative admissible sources and duplicate candidates. It can be an explicitly justified premise or a supported collection/retention statement. This paper specifies the needed basis, without designing its implementation. For positive coverage of \(q\), an admissible satisfying record with supported membership in \(V_W(t_c)\) suffices; exhaustive \(\mathrm{ClosedView}(q,W,t_c)\) is not additionally required for that existential claim.

Three absence claims must be distinguished:

\begin{center}
\fontsize{8}{9.6}\selectfont
\setlength{\tabcolsep}{2pt}
\renewcommand{\arraystretch}{1.10}
\begin{tabularx}{\columnwidth}{@{}L{0.25\columnwidth} L{0.35\columnwidth} Y@{}}
\toprule
\textbf{Claim} & \textbf{Basis} & \textbf{Consequence} \\
\midrule
Retrieved-set absence & No satisfying record was found in the current retrieval & May leave coverage unresolved; another relevant record could be available but unretrieved \\
\addlinespace[3pt]
Verifier-view absence & Relevant view closure is justified, and every candidate is conclusively inadmissible or no candidate exists & Supports a conclusively uncovered obligation at \(t_c\) \\
\addlinespace[3pt]
Historical non-receipt by \(d_q\) & Adequate first-receipt history or retention history excludes any admissible satisfying record reaching \(W\) by \(d_q\) & Supports a historical deadline miss once the deadline has passed \\
\bottomrule
\end{tabularx}
\end{center}

An obligation is \textbf{conclusively uncovered} when it is known applicable and due, relevant view closure holds, and no record in that view satisfies its admissibility predicate. An unresolved potentially admissible candidate prevents that negative conclusion. A wrong-bound record fails the duty, but all possible satisfying candidates in the relevant view must still be considered.

Current absence does not entail historical non-delivery. For example, a record received on time may subsequently leave a limited-retention view. Conversely, a closed current view combined with a supported no-removal retention history can establish a deadline miss when that history would have retained every earlier admissible arrival.

\subsection{Aggregate result and assessment procedure}
\label{sec:4.3}

The reporting rule has the following precedence.

\begin{table*}[t]
\caption{Coverage results at the cutoff.}
\label{tab:coverage}
\centering
\fontsize{8}{9.6}\selectfont
\setlength{\tabcolsep}{2pt}
\renewcommand{\arraystretch}{1.12}
\begin{tabularx}{\textwidth}{@{}L{0.085\textwidth} L{0.29\textwidth} Y@{}}
\toprule
\textbf{Precedence} & \textbf{Result} & \textbf{Required basis} \\
\midrule
1 & \texttt{INCOMPLETE} & At least one known applicable, due instance is conclusively uncovered in the relevant closed verifier view. Name each supported gap. Other unknowns remain diagnostics. \\
\addlinespace[3pt]
2 & \texttt{COMPLETE\_\allowbreak{}WITHIN\_\allowbreak{}SCOPE} & The scope is fixed, \(\mathrm{BranchJustified}(b,S)\) and \(\mathrm{ClosedObligations}(S)\) hold, every applicable due instance has an admissible and correctly bound satisfying record established to belong to \(V_W(t_c)\), and any additional closure premise actually used is stated and supported. \\
\addlinespace[3pt]
3 & \texttt{UNKNOWN} & Neither a supported specific gap nor positive completeness can be established. Identify the unresolved branch, inventory, view, time, or admissibility basis. \\
\bottomrule
\end{tabularx}
\end{table*}

Obligation closure concerns the completeness of the applicable inventory. Positive coverage concerns admissible satisfying records and their supported membership in \(V_W(t_c)\); it does not universally require exhaustive closure of that view. Additional closure premises actually used to justify the branch or inventory must still be supported. Any part of the conclusion that relies on absence needs the corresponding view-closure basis.

A reusable assessment proceeds as follows:

\begin{enumerate}
\item Fix \(S\), the profile version, clock basis, and assessment premises.
\item Identify the basis for \(\mathrm{BranchJustified}(b,S)\) and the triggers; generate the obligation inventory, recording its closure basis and unresolved parts.
\item Separate applicable due instances from known inapplicable and not-yet-due instances.
\item For each due instance, identify admissible records or a supported absence; preserve unresolved candidate and view checks.
\item Apply Table~\ref{tab:coverage} and report the scope, branch, inventory basis, record-to-obligation mapping, specific gaps, unresolved checks, and supporting premises.
\end{enumerate}

This procedure does not select the branch with fewer obligations when the branch basis is uncertain. If no specific gap is established, unresolved branch or inventory closure yields \texttt{UNKNOWN}, even when every item on a provisional list has a record.

\subsection{Separate progress, outcome knowledge, and timeliness}
\label{sec:4.4}

A coverage result is accompanied by three distinct descriptive outputs:

\begin{itemize}
\item \textbf{Execution progress:} the stages supported by the available stage records under the declared premises.
\item \textbf{External outcome knowledge:} the outcome information available to \(W\), identifying its source and assertion limits; it may remain \texttt{UNKNOWN}.
\item \textbf{Historical delivery timeliness:} for each triggered duty, whether an admissible satisfying record is known to have reached \(W\) by \(d_q\), is known not to have done so after the deadline, or has unresolved delivery history.
\end{itemize}

Timeliness concerns any admissible satisfying record, not merely the arrival time of one selected copy. A miss therefore needs a basis excluding other timely satisfying records. If admissibility includes independent temporal qualifications, a historical assessment must also justify those qualifications for the historical claim.

\mbox{\texttt{COMPLETE\_WITHIN\_SCOPE}} covers obligations due \textbf{so far} within \(H\); later due duties and untriggered stages must remain visible in the report. It does not announce lifecycle termination or successful external execution. A final failure report can cover a final-result duty if it meets the profile's predicate. \texttt{INCOMPLETE} and \texttt{UNKNOWN} do not establish failure or non-occurrence. A later admissible arrival can change current coverage while leaving the earlier cutoff assessment and a supported deadline miss intact.

\section{Constructed Execution Profile}
\label{sec:5}

\subsection{Scope and path}
\label{sec:5.1}

Consider Intent \(I\) to disable service principal \texttt{sp-17} in tenant \texttt{T}. Candidate \(C\) is the exact API request setting that principal's \texttt{enabled} field to \texttt{false}, including tenant, target, method, endpoint, request body, and request identifier. \(A\) selects one admitted attempt for this Candidate. Boundary \(B\) includes the Intent/admission source, Candidate materializer, adjudicator, execution controller, Executor, status reporter, external identity-service result source, and verifier \(W\).

The fixed profile has dispatch and terminal-refusal branches. The declared horizon includes the selected branch's terminal reporting stage: final-result reporting after invocation, or terminal-refusal reporting. For this bounded construction, \(P\) stipulates at most one external invocation for the selected \(A\); internal retries are excluded. This bound is a profile premise, not an inference from finding one invocation record.

The dispatch path is:

Intent establishment \(\rightarrow\) Candidate materialization \(\rightarrow\) adjudication \(\rightarrow\) local dispatch commit \(\rightarrow\) external invocation \(\rightarrow\) scheduled status report \(\rightarrow\) final-result record.

A \textbf{local dispatch commit} is the durable local event that the execution controller accepted the exact Candidate for dispatch under this constructed profile. Its record attests that local acceptance. External invocation, external acceptance, final state change, and business success require separate information. The controller is a control-plane component; no hardware device is assumed.

Adjudication selects the branch in this construction. Its retrospective record is not a Paper-3 execution-permission state. The profile makes no ERC, Grant, or Redemption claim.

\subsection{Instantiated record obligations}
\label{sec:5.2}

Table~\ref{tab:profile} defines eight obligation types. \(q_I,\ldots,q_R\) label their selected instances; \(I_r,\ldots,R_r\) label candidate records. All accepted records must pass the declared integrity and source-qualification checks. We use \textbf{final-result record} (\(R_r\)) for the record containing the designated external result source's result assertion or Receipt.

\begin{table*}[t]
\caption{Constructed branch-specific profile. The source assignment and record predicates are analytical stipulations.}
\label{tab:profile}
\centering
\fontsize{8}{9.6}\selectfont
\setlength{\tabcolsep}{2pt}
\renewcommand{\arraystretch}{1.12}
\begin{tabularx}{\textwidth}{@{}L{0.20\textwidth} L{0.27\textwidth} Y L{0.18\textwidth}@{}}
\toprule
\textbf{Obligation and activating trigger} & \textbf{Required content and admissible source} & \textbf{Binding and stage predicates} & \textbf{Branch; availability at W} \\
\midrule
\(q_I\): admission of \(A\) for \(I\) & \(I_r\): structured Intent and admitted selection, from the Intent/admission source & Identifies \(I,A\); supplies the Intent used in adjudication & Both; \(d_I\) \\
\addlinespace[3pt]
\(q_C\): materialization of \(C\) & \(C_r\): exact Candidate and request identifier, from the Candidate materializer & Binds \(C\) to \(I,A\); supplies the Candidate used in adjudication & Both; \(d_C\) \\
\addlinespace[3pt]
\(q_J\): adjudication & \(J_r\): branch decision and referenced input basis, from the adjudicator & Identifies \(I,C,A\) and the preceding Intent/Candidate records & Both; \(d_J\) \\
\addlinespace[3pt]
\(q_N\): selection of the refusal branch & \(N_r\): terminal local refusal disposition and reason, from the execution controller & Binds \(I,C,A,J_r\); records closure of the selected local attempt under the refusal path & Refusal; \(d_N\) \\
\addlinespace[3pt]
\(q_K\): local dispatch commit & \(K_r\): durable local dispatch acceptance, from the execution controller & Binds \(I,C,A,J_r\); follows dispatch-branch adjudication & Dispatch, when commit occurs; \(d_K\) \\
\addlinespace[3pt]
\(q_X\): external invocation & \(X_r\): exact request submitted and its request identifier, from the Executor & Binds \(I,C,A,K_r\); follows local dispatch commit & Dispatch, when invocation occurs; \(d_X\) \\
\addlinespace[3pt]
\(q_U\): invocation schedules one status report & \(U_r\): status at the scheduled check, including \texttt{OUTCOME\_\allowbreak{}UNKNOWN} when appropriate, from the designated status reporter & Binds the external request through \(X_r\); identifies the check time and the reporter's knowledge & Invoked dispatch; \(d_U\) \\
\addlinespace[3pt]
\(q_R\): invocation activates final-result reporting & \(R_r\): final-result record, from the designated identity-service result source & Identifies the external request, tenant, target, and final result; verified correlation through \(X_r,C_r\) binds it to \(C,A,I\); creation follows the asserted observation & Invoked dispatch; \(d_R\), fixed at invocation in this constructed \(P\) \\
\bottomrule
\end{tabularx}
\end{table*}

Under this constructed \(P\), invocation activates the final-result duty. \(R_r\) can be produced only after the final event it describes, but its availability deadline does not wait for that event. If no final result exists by the deadline, the required final-result record can remain uncovered. Similarly, invocation schedules the status duty; failing to perform the scheduled check does not eliminate that duty. A truthful \texttt{OUTCOME\_\allowbreak{}UNKNOWN} report can satisfy \(q_U\) and cannot substitute for \(q_R\).

The fixed final-result deadline is a property of this constructed profile, not a universal requirement of evidence coverage. An unmet duty means that this profile's required record is not covered by the deadline; it does not establish execution failure, logging-system failure, or external non-occurrence. A different profile could use repeated status obligations, a later terminalization rule, or no fixed final-result deadline. The applicable reporting rule must be declared before assessment, rather than selected after the observed evidence is known.

The external result source attests its own request and result information. Internal attempt and predecessor bindings can be checked through the stipulated request correlation; the source does not attest the controller's local commit.

Under a justified terminal-refusal branch, the inventory is \(q_I,q_C,q_J,q_N\). \(J_r\) records adjudication; \(N_r\) records the controller's terminal disposition, so they have distinct predicates. No dispatch, invocation, status, or final-result trigger occurs for the selected attempt under this profile premise. The terminal-refusal record supplies part of the branch-closure basis; its presence alone does not prove the absence of all external actions through other paths or attempts.

For the invoked dispatch branch, the inventory is \(q_I,\allowbreak{}q_C,\allowbreak{}q_J,\allowbreak{}q_K,\allowbreak{}q_X,\allowbreak{}q_U,\allowbreak{}q_R\). Establishing that inventory requires grounds for the actual triggers and the one-invocation bound. In a less resolved execution, branch selection alone would not establish all later triggers or their absence.

\section{Worked Analytical Cases}
\label{sec:6}

\subsection{Shared construction and Case A: complete selected branches}
\label{sec:6.1}

All times are constructed analytical data on a common observer clock. They are not production measurements. The profile sets

\[
\begin{aligned}
&d_I=2,\quad d_C=3,\quad d_J=5,\quad d_N=6,\\
&d_K=8,\quad d_X=9,\quad d_U=15,\quad d_R=30.
\end{aligned}
\]

For the dispatch path, Intent admission occurs at 0, Candidate materialization at 1, adjudication at 4, local dispatch commit at 6, and invocation at 7. The corresponding records are created at those event times. First receipt at \(W\) is 1 for \(I_r\), 2 for \(C_r\), 5 for \(J_r\), 7 for \(K_r\), and 8 for \(X_r\). The scheduled status check and record creation occur at 12; \(W\) receives \(U_r\) at 12, with \texttt{OUTCOME\_\allowbreak{}UNKNOWN} as the status reporter's knowledge.

The cases stipulate competent sources, verified bindings, and an adequate branch and trigger-enumeration basis within the selected scope. From admission at time 0 through time 41, \(W\) retains every received relevant record, together with trustworthy first-receipt information, without removal. Inspection exhausts this retained relevant view, including all possible satisfying candidates. These stipulations supply obligation closure and cutoff-view membership for positive coverage, as well as relevant view closure for absence findings and a historical delivery basis. They are not conclusions derived from the coverage labels.

\textbf{Case A, dispatch.} The designated external source observes the final result at 24 and creates its record at 25; \(W\) receives \(R_r\) at 25. At \(t_c=31\), the closed inventory comprises the seven dispatch duties, all due and admissibly covered. The result is \mbox{\texttt{COMPLETE\_WITHIN\_SCOPE}}. Progress includes invocation and final-result reporting. Outcome information consists of the external source's reported final disablement for the selected request. All specified availability deadlines were met. The earlier unknown status at 12 is compatible with a final report concerning time 24.

\textbf{Case A, refusal.} Replace the dispatch branch with refusal adjudication at 4 and terminal local refusal at 5. \(W\) receives \(J_r\) at 5 and \(N_r\) at 6, together with the same timely Intent and Candidate records. At 31, the justified inventory comprises \(q_I,q_C,q_J,q_N\), and coverage is \mbox{\texttt{COMPLETE\_WITHIN\_SCOPE}}. Progress ends at terminal local refusal. No external outcome is asserted for an invocation on this selected branch. All four applicable delivery deadlines were met. Missing later-stage record types are not gaps because their triggers are inapplicable under the justified terminal branch.

\subsection{Case B: a due final-result record missing, then delivered late}
\label{sec:6.2}

Case B uses the same invoked dispatch scope and seven-instance inventory. It is the principal walkthrough of the assessment method. The early records and status report arrive as above. \(W\) has no other outcome information at time 31. The external source's final-result record is first received at time 40; it describes an observation at 24 and is created at 25. The construction specifies that later record for comparison, while the time-31 assessment uses only information available to \(W\) then.

\begin{table*}[t]
\caption{Case B evidence ledger. ``Admissible'' applies when the indicated record is available. The historical column uses the stipulated complete retention and first-receipt history, excluding another timely satisfying record.}
\label{tab:ledger}
\centering
\fontsize{8}{9.6}\selectfont
\setlength{\tabcolsep}{2pt}
\renewcommand{\arraystretch}{1.12}
\begin{tabularx}{\textwidth}{@{}L{0.074\textwidth} L{0.159\textwidth} L{0.050\textwidth} L{0.209\textwidth} L{0.058\textwidth} L{0.100\textwidth} L{0.070\textwidth} L{0.070\textwidth} Y@{}}
\toprule
\textbf{Obligation} & \textbf{Triggered/applicable?} & \textbf{Due at W} & \textbf{Record candidate} & \textbf{First receipt} & \textbf{Admissible?} & \textbf{Covered at 31?} & \textbf{Covered at 41?} & \textbf{Historical delivery timeliness} \\
\midrule
\(q_I\) & Yes; admission at 0, both branches & 2 & \(I_r\) & 1 & Yes & Yes & Yes & Met: \(1\leq2\) \\
\addlinespace[3pt]
\(q_C\) & Yes; materialization at 1, both branches & 3 & \(C_r\) & 2 & Yes & Yes & Yes & Met: \(2\leq3\) \\
\addlinespace[3pt]
\(q_J\) & Yes; adjudication at 4, dispatch selected & 5 & \(J_r\) & 5 & Yes & Yes & Yes & Met: \(5\leq5\) \\
\addlinespace[3pt]
\(q_K\) & Yes; local dispatch commit at 6 & 8 & \(K_r\) & 7 & Yes & Yes & Yes & Met: \(7\leq8\) \\
\addlinespace[3pt]
\(q_X\) & Yes; invocation at 7 & 9 & \(X_r\) & 8 & Yes & Yes & Yes & Met: \(8\leq9\) \\
\addlinespace[3pt]
\(q_U\) & Yes; invocation schedules the check at 12 & 15 & \(U_r\): \texttt{OUTCOME\_\allowbreak{}UNKNOWN} at 12 & 12 & Yes & Yes & Yes & Met: \(12\leq15\) \\
\addlinespace[3pt]
\(q_R\) & Yes; final-result duty activated at invocation 7 & 30 & None available at 31; \(R_r\) arrives at 40, with event at 24 and creation at 25 & 40 & Yes upon receipt; no candidate at 31 & No, conclusively uncovered & Yes & Missed: no admissible satisfying arrival by 30; first such receipt is 40 \\
\bottomrule
\end{tabularx}
\end{table*}

\(q_N\) is known inapplicable on this branch, so it is excluded from the due ledger. All seven listed instances are due at both assessment cutoffs.

\textbf{Step 1: establish the scope and inventory.} \(I,C,A,B,H,P,W\) are fixed. Branch and trigger premises justify the invoked dispatch inventory, including \(q_R\). Its deadline was fixed when invocation activated the duty; it is not conditional on \(W\) having already received \(R_r\).

\textbf{Step 2: assess admissible records at 31.} The first six duties have admissible satisfying records. In particular, \(U_r\) arrived at 12 before \(d_U=15\). Its unknown outcome content meets the status predicate. It says what the status reporter knew at the check; it does not meet the separate final-result predicate.

\textbf{Step 3: establish the specific gap.} \(q_R\) is known applicable and \(30\leq31\). Exhaustive inspection of the relevant retained view finds no admissible \(R_r\). View closure therefore makes this a conclusive coverage gap rather than mere retrieved-set absence.

\textbf{Step 4: report the time-31 result.} Table~\ref{tab:coverage} gives \texttt{INCOMPLETE}, identifying \(q_R\). Progress is supported through invocation and the scheduled status check. External outcome knowledge remains \texttt{UNKNOWN}. The retention and first-receipt premises additionally exclude an admissible arrival by 30, supporting a historical deadline miss. That timing finding requires the historical premises; current absence alone would not establish it.

\textbf{Step 5: reassess at 41.} At 40, \(W\) first obtains an admissible \(R_r\), bound to the same request through \(X_r,C_r\). In \(V_W(41)\), every applicable due duty is covered, so the new result is \mbox{\texttt{COMPLETE\_WITHIN\_SCOPE}}. Progress now includes final-result reporting, and \(W\) has the external source's final-disablement assertion. The historical delivery finding remains a miss because \(40>30\).

The event timestamp 24 and record creation time 25 do not establish receipt by 30. The later assessment uses a later view. It leaves the time-31 \texttt{INCOMPLETE} finding correct for its cutoff and preserves the supported deadline miss.

\subsection{Case C: covered selected attempt, unresolved all-attempt inventory}
\label{sec:6.3}

Let \(A_1\) have Case A's complete dispatch account at time 31. Its explicitly selected-attempt scope is \mbox{\texttt{COMPLETE\_WITHIN\_SCOPE}}. Progress and outcome information are those of Case A, and its seven availability deadlines were met.

A broader proposed claim asks whether every admitted attempt associated with \(I\) within a stated boundary and the interval from 0 through 31 has been covered. This requires an adequate attempt-enumeration source and a justified end or closure condition for that enumeration. Each enumerated attempt also needs its own Candidate, branch, horizon, and generated duties with their deadline instantiation on the declared clock basis. The claim is an assessment over a family of scopes; sharing \(I\) does not collapse their Candidates or duties into \(A_1\)'s inventory.

Suppose \(W\) has neither a supported complete enumeration nor an adequate end condition for that broader interval. No additional known applicable, due duty is conclusively uncovered on the available information. The broader result is \texttt{UNKNOWN}, because obligation-inventory closure is unresolved. The known progress, outcome report, and timely deliveries for \(A_1\) remain reportable; aggregate progress, outcome information, and timeliness across the unenumerated attempts remain unresolved.

If a further attempt \(A_2\) is established, its duties must be examined. A supported specific due gap for \(A_2\) would yield \texttt{INCOMPLETE} for the broader assessment even if other enumeration questions remained. Its mere existence does not invalidate the earlier explicitly selected-\(A_1\) conclusion. This analysis states the premises needed to enlarge a coverage claim; it introduces no business-operation identity or retry protocol.

\section{Discussion and Limitations}
\label{sec:7}

\subsection{Conditional completeness and the reference basis}
\label{sec:7.1}

\mbox{\texttt{COMPLETE\_WITHIN\_SCOPE}} is conditional on the declared profile, justified branch, obligation closure, source/admissibility premises, supported membership of the satisfying records in the cutoff view, and any additional closure premises actually used. A profile that omits a useful stage can be completely covered according to its own duties. The label therefore says what that profile requires and the evidence covers; it does not certify that the profile is sufficient for every audit purpose. Profile adequacy must be assessed for the intended application, consistently with the application-relative concern in \cite{souilah2009}.

The constructed cases stipulate their closure and source premises to make the assessment inspectable. A deployment assessment would need to justify them, especially the event/instance enumeration that makes the inventory closed and the retention/indexing basis that makes absence conclusive. The model provides no universal algorithm for discovering hidden triggers or obligations, and no deployment-completeness certificate.

\subsection{Boundaries of the conclusion}
\label{sec:7.2}

\textbf{Coverage and truth.} An intact, scope-complete record set can contain a competent source's mistaken assertion. The method establishes record coverage under its premises. It does not define a general record-to-fact support relation, competing external-truth interpretations, or revision of source trust.

\textbf{Coverage and authorization.} The labels do not produce \texttt{ALLOW}, an ERC, a Grant, or Redemption permission. The retrospective duties defined here do not modify the pre-execution Evidence Obligations or other conformance conditions of \cite{p2}.

\textbf{Coverage and causal authority.} A gap does not establish bypass or identify who could cause an action. A terminal-refusal account is conditional on its scope and branch premises. Causal realization, veto coverage, and trust-domain independence require the analysis of \cite{p1}.

\textbf{Coverage and recovery.} The report does not decide whether to retry, compensate, freeze, or recover after uncertainty. Nor does a later arrival erase an earlier view or implement a general judgment-revision model.

\subsection{Diagnostic use and extensibility}
\label{sec:7.3}

A missing duty can motivate investigation of observation, record generation, preservation, or exposure. An event might lie outside effective observation; an observed event might lack a generated record; a generated record might be lost; or a retained record might not be exposed to \(W\). These possibilities are neither exhaustive nor mutually exclusive. Absence alone does not identify the cause.

Other deployments can instantiate different stages, schemas, qualified sources, closure artifacts, and verifier architectures. Logging and provenance representations can carry the relevant records and closure information. Such an instantiation must state its own trigger rules, deadlines, binding predicates, and assessment premises. The cases illustrate consistent use of the method; they provide no empirical evaluation, performance result, or proof that a real control plane meets those premises.

\section{Conclusion}
\label{sec:8}

Retrospective execution coverage requires a declared object of assessment and a justified account of the records due for it. We have defined a scope-bound model and a reporting method that combines branch and trigger applicability, a stage horizon, a closed obligation inventory, admissible source and object bindings, a named verifier view, and a cutoff. Constructed dispatch, refusal, late-receipt, and all-attempt cases make the three result labels inspectable. The method permits a later view to close a current gap while preserving an earlier deadline miss. Its completeness conclusion remains conditional on the profile and closure premises, with execution progress, outcome knowledge, and historical timeliness reported separately.

\balance

\end{document}